\documentclass[10pt,twocolumn,letterpaper]{article}

\usepackage{iccv}
\usepackage{times}
\usepackage{epsfig}
\usepackage{graphicx}
\usepackage{amsmath}
\usepackage{amssymb}
\usepackage{enumitem}
\usepackage{multirow}
\usepackage[accsupp]{axessibility}
\newcommand\norm[1]{\left\lVert#1\right\rVert}

\usepackage[breaklinks=true,bookmarks=false]{hyperref}

\iccvfinalcopy 

\def\iccvPaperID{8243} 

\ificcvfinal\fi

\begin{document}

\title{Learning Conditional Knowledge Distillation for Degraded-Reference \\ Image Quality Assessment}

\author{Heliang Zheng$^{1}\thanks{This work was performed when Heliang Zheng was visiting Microsoft Research as a research intern. Code and models are available at \url{https://github.com/researchmm/CKDN}.}$  , Huan Yang$^{2}$, Jianlong Fu$^{2}$, Zheng-Jun Zha$^{1}$, Jiebo Luo$^{3}$
\\ \small $^1$University of Science and Technology of China, Hefei, China
\\ \small $^2$Microsoft Research, Beijing, China
\\ \small $^3$University of Rochester, Rochester, NY
\\ \small zhenghl@mail.ustc.edu.cn, \{huayan, jianf\}@microsoft.com, zhazj@ustc.edu.cn, jluo@cs.rochester.edu
}

\maketitle
\ificcvfinal\thispagestyle{empty}\fi

\begin{abstract}

An important scenario for image quality assessment (IQA) is to evaluate image restoration (IR) algorithms. The state-of-the-art approaches adopt a full-reference paradigm that compares restored images with their corresponding pristine-quality images. However, pristine-quality images are usually unavailable in blind image restoration tasks and real-world scenarios. In this paper, we propose a practical solution named degraded-reference IQA (DR-IQA), which exploits the inputs of IR models, degraded images, as references. Specifically, we extract reference information from degraded images by distilling knowledge from pristine-quality images. The distillation is achieved through learning a reference space, where various degraded images are encouraged to share the same feature statistics with pristine-quality images. And the reference space is optimized to capture deep image priors that are useful for quality assessment. Note that pristine-quality images are only used during training. Our work provides a powerful and differentiable metric for blind IRs, especially for GAN-based methods. Extensive experiments show that our results can even be close to the performance of full-reference settings.

\end{abstract}

\vspace{-3mm}

\section{Introduction}

\begin{figure}[t]
    \centering
    \includegraphics[width=\linewidth]{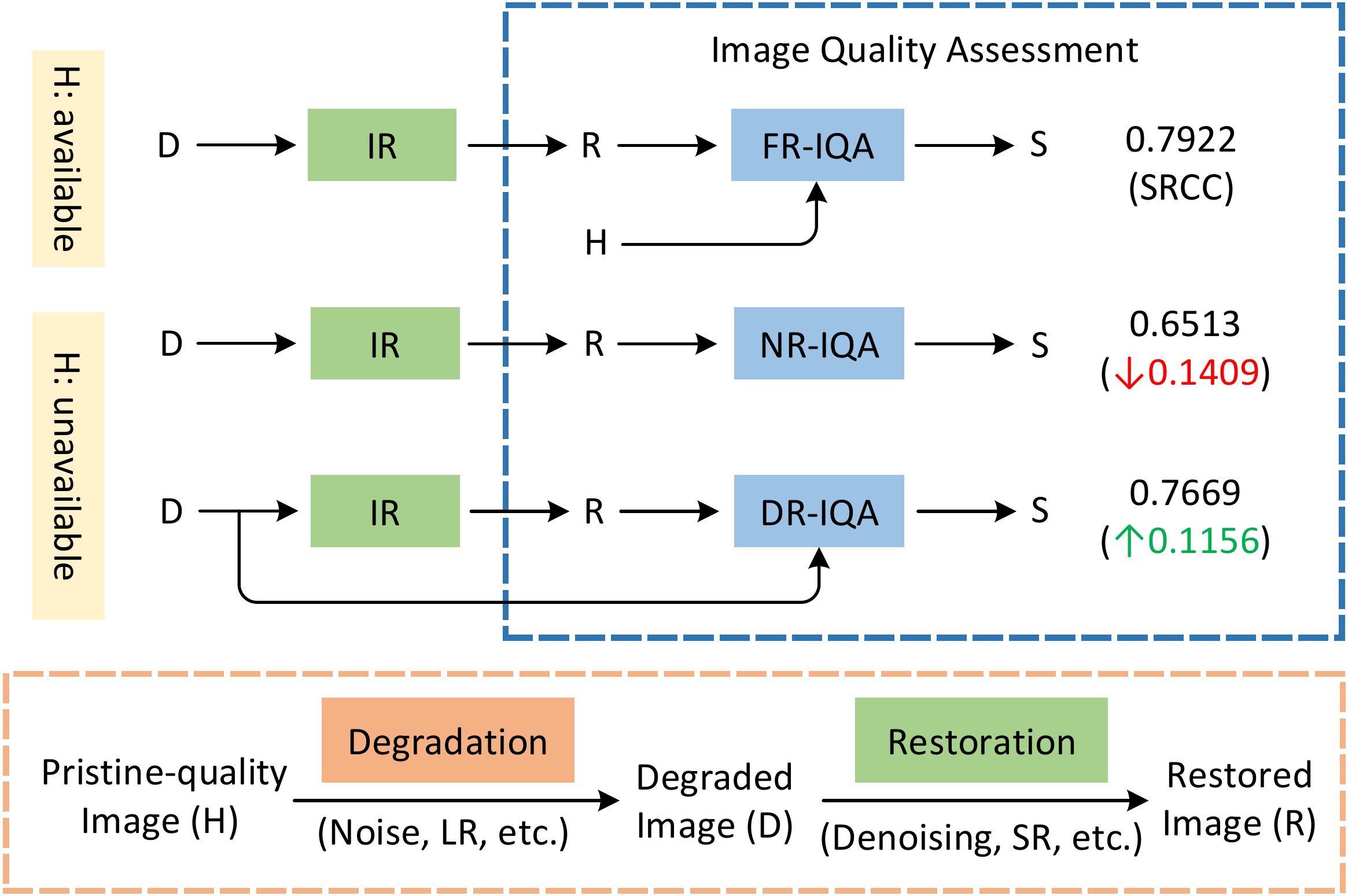}
    \vspace{-2mm}
    \caption{(a) Pristine-quality images can provide strong reference information for IQA. When pristine-quality images are unavailable, (b) directly regressing restored images to quality scores causes a dramatic drop in performance. To this end, (c) we propose to extract reference information from degraded images and make such a solution effective. FR, NR, DR, S, and SRCC indicate full-reference, no-reference, degraded-reference, quality score, and Spearman's Rank order Correlation Coefficients~\cite{sheikh2006statistical}, respectively.}
    \label{fig:teaser}
    \vspace{-4mm}
\end{figure}

Digital images may be subject to various quality degradations during processing, compression, transmission, etc.~\cite{wang2004image}. And image restoration (IR) algorithms are developed to improve the quality of degraded images~\cite{ma2020structure,wang2018esrgan,yang2020learning,zeng2019learning,zhang2017beyond,zhang2019ranksrgan,zhang2018image}. A consequential question is how to assess the quality of restored images and evaluate IR algorithms. The state-of-the-art IQA approaches adopt a full-reference paradigm (FR-IQA), which learns to compare restored images with their corresponding pristine-quality images (as shown in Figure~\ref{fig:teaser} (a)). FR-IQA algorithms have been widely adopted as IR evaluation metrics, e.g., PSNR, SSIM~\cite{wang2004image}, and LPIPS~\cite{zhang2018unreasonable}. However, they cannot be applied to blind image restoration tasks and real-world applications, where pristine-quality images are unavailable. In this paper, we study the problem of evaluating IR models without relying on pristine-quality images.

To evaluate IR models without pristine-quality images, no-reference IQA (NR-IQA) methods provide a solution, that is, directly regressing restored images to quality scores, i.e., mean opinion scores (MOS)~\cite{bosse2017deep,kang2014convolutional,ma2017end,talebi2018nima}. However, the absence of reference information makes the problem much more challenging and causes a dramatic drop in performance, e.g., $0.1409$ SRCC (Spearman's Rank order Correlation Coefficients~\cite{sheikh2006statistical}) drops. This motivates us to seek available reference information. We find that the inputs of IR models (i.e., degraded images) are usually free to be obtained, and it has been verified that degraded images also contain useful image priors like pristine-quality images for solving under-constrained problems in computer vision~\cite{pan2020exploiting,ulyanov2018deep}. To this end, we propose a new solution named degraded-reference IQA (DR-IQA). However, directly replacing the pristine-quality references of existing FR-IQA models~\cite{jinjin2020pipal} by degraded references causes 0.1239 SRCC drops, because the noises involved by various degradations make it difficult to mine and leverage reference information from degraded images.

\begin{figure*}[!t]
    \centering
    \includegraphics[width=.98\linewidth]{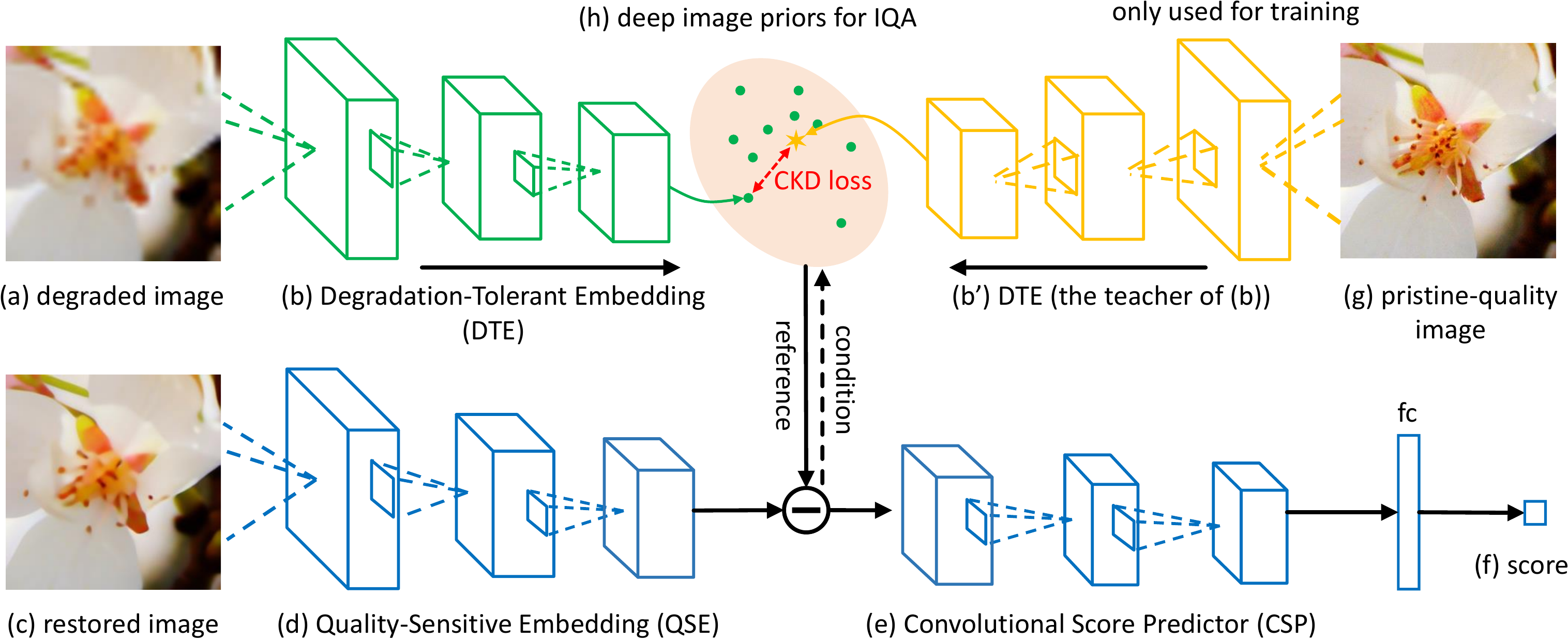}
    \caption{Overview of the proposed Conditional Knowledge Distillation Network (CKDN). CKDN consists of three components, i.e., a degradation-tolerant embedding module (DTE) in (b), a quality-sensitive embedding module (QSE) in (d), and a convolutional score predictor (CSP) in (e). The DTE embeds the degraded image in (a) to a reference space in (h) by optimizing a conditional knowledge distillation (CKD) loss. Such a loss learns 
   reference information from pristine-quality images in (g) and minimizes the difference between the features of degraded and pristine-quality images. Note that the QSE and CSP provide conditions to learn the reference space, and the learned representation of degraded images provides reference information for IQA.}
    \vspace{-3mm}
    \label{fig:overview}
\end{figure*}

To leverage degraded images and improve the effectiveness of DR-IQA, we propose a \textbf{C}onditional \textbf{K}nowledge \textbf{D}istillation \textbf{N}etwork (CKDN). As shown in Figure~\ref{fig:overview}, CKDN consists of three modules, i.e., a degradation-tolerant embedding module (DTE), a quality-sensitive embedding module (QSE), and a convolutional score predictor (CSP). The DTE is the key module, which aims to effectively extract reference information for IQA from degraded images. Specifically, it learns a reference space, where various degraded images are optimized to share the same feature statistics with pristine-quality images. Such space is learned in the condition of conducting quality assessments with the QSE and CSP, where the QSE learns discriminative features from restored images, and the CSP maps feature differences to human-annotated scores. Moreover, we find that pre-training such conditions can help to learn quality-sensitive features and further advance the optimization of the reference space. In particular, we propose a relative score regression task that can enlarge the space of training data by creating data pairs.

We conduct extensive experiments to evaluate our proposed solution. It is shown that our proposed CKDN enables DR-IQA to achieve comparable performance to the full-reference setting on various IR tasks, e.g., traditional/GAN-based super-resolution, de-noising, etc. Moreover, we further study the influence of reference quality and the performance of current IQA methods on evaluating different IR tasks. We draw three conclusive insights: 1) our CKDN works well with a large range of degradation types, 2) reference images are extremely important for evaluating GAN-based images, and 3) current IQA methods can provide around 85\% reliable judgments for evaluating IR algorithms. Please refer to Section~\ref{sec:dis} for more details. Our main contributions can be summarized as follows: 1) To the best of our knowledge, we are the first to leverage degraded images for IQA. 2) We formulate this practical setting and propose an effective conditional knowledge distillation network to solve this problem. 3) Extensive analyses show that degraded references are of great help for evaluating GAN-based images. Overall, we believe our work would contribute to the IQA and the blind SR communities, and providing insights for evaluating GAN-based models.

\vspace{-2mm}
\section{Related Work}
\label{sec:rw}
\vspace{-2mm}
\textbf{No-reference Image Quality Assessment.} NR-IQA aims to predict the perceived quality of distorted images without referring to pristine-quality images. Getting rid of reference images makes such a setting more practical in many applications and therefore receives more and more attention in recent years. Early works in the literature of NR-IQA proposed to model statistics of natural images and regress parametric deviations to image degradations~\cite{mittal2012no,mittal2012making,saad2012blind}. In recent years, deep neural networks (DNNs) show promising capacity and are leveraged to fit distorted images to quality scores for NR-IQA~\cite{bosse2017deep,kang2014convolutional,ma2017end,talebi2018nima}. Considering the problem of limited training data, dipIQ proposed to learn-to-rank large-scale unlabeled image pairs in a self-supervised manner ~\cite{ma2017dipiq}. Inspired by FR-IQA, Hallucinated-IQA proposed to compare the distorted images to hallucinated references, which are restored from the distorted images by adversarial training~\cite{chen2020no,lin2018hallucinated,ren2018ran4iqa}. And some researchers also propose to generate a quality map for NR-IQA ~\cite{pan2018blind}. There are three differences compared to these works: 1) Compared to hallucinated references that generated from assessed images, degraded references are independent of assessed images and can provide additional information. 2) Previous works use a handcraft space (i.e., RGB image/SSIM map), while we learn deep embeddings end-to-end. 3) previous works try to ``distill'' all the information instead of only reference information for IQA.

\textbf{Full-reference Image Quality Assessment.} 
The full-reference setting significantly simplifies the IQA problem by converting it to two sub-problems, i.e., calculating the distance between restored and reference images and mapping such distance to quality scores~\cite{mantiuk2011hdr,wang2004image,wang2003multiscale,zhang2011fsim}. Here we mainly review the full-reference IQA methods that have been widely used to evaluate IR models~\cite{wang2004image,zhang2018unreasonable}. PSNR and SSIM~\cite{wang2004image} are appealing because they are simple to calculate and have clear physical meanings. But they do not correlate well with perceived visual quality~\cite{jinjin2020pipal,prashnani2018pieapp,zhang2018unreasonable}, especially for evaluating the images generated by GANs~\cite{goodfellow2014generative}. LPIPS~\cite{zhang2018unreasonable} analyzed the unreasonable effectiveness of deep features as a perceptual metric, which has been adopted by more and more IR models as a perceptual metric~\cite{gu2020interpreting,ma2020structure}. And SWDN further improves LPIPS by taking the spatial misalignment for GAN-based distortion into consideration~\cite{jinjin2020pipal,gu2020image}. Different from FR-IQA, our proposed DR-IQA only requires degraded images as reference, and we proposed a Conditional Knowledge Distillation Network to leverage deep image priors from the degraded images. Note that our proposed DR-IQA is also different from traditional reduced-reference IQA (RR-IQA), which extracts predefined features, such as frequency domain coefficients and image gradients, as references under the scenario of visual communication systems~\cite{li2009reduced,wang2013reduced}.

\vspace{-2mm}
\section{Approach}
\vspace{-1mm}

\begin{figure}[!t]
    \centering
    \includegraphics[width=.85\linewidth]{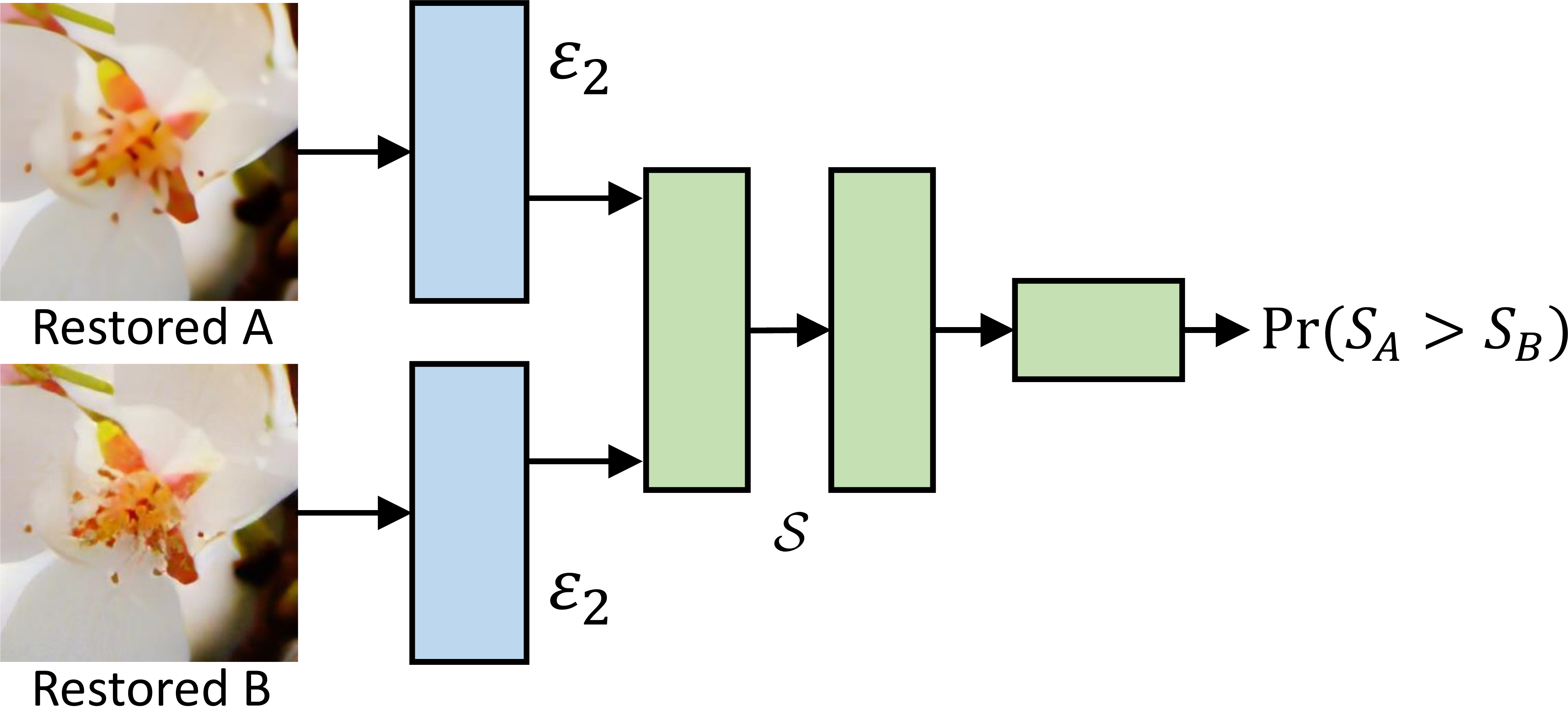}
    \vspace{2mm}
    \caption{An illustration of pre-training the QSE by the proposed relative score regression loss.}
    \vspace{-2mm}
    \label{fig:pre}
\end{figure}

\label{sec:method}
In this section, we will introduce the setting of DR-IQA and our proposed model for DR-IQA. Our model takes as input a degraded image in Figure~\ref{fig:overview} (a) (i.e., the input of IR models) and a restored image in Figure~\ref{fig:overview} (c) (i.e., the output of IR models), and aims to predict a quality score for the restored image in Figure~\ref{fig:overview} (f). Similar to FR-IQA, the basic idea is to compare the restored image to the reference and further map the learned difference to a quality score. However, the reference images in DR-IQA are noisy due to degradations. To solve this problem, we propose to learn a reference space in Figure~\ref{fig:overview} (h) to capture deep image priors that are useful for quality assessment. In particular, our proposed  Conditional Knowledge Distillation Network (CKDN) consists of a degradation-tolerant embedding module (DTE) in Figure~\ref{fig:overview} (b), a quality-sensitive embedding module (QSE) in Figure~\ref{fig:overview} (d), and a convolutional score predictor (CSP) in Figure~\ref{fig:overview} (e). We propose a conditional knowledge distillation (CKD) loss to guide the learning of the reference space. Moreover, we propose a relative score regression loss to pre-train the QSE (shown in Figure~\ref{fig:pre}). We will introduce the details in the following.  

\vspace{-1mm}
\subsection{Formulations}
\vspace{-2mm}
\textbf{Formulation of DR-IQA.} Given a pristine-quality image $\mathbf{H}$; the degraded image can be denoted as $\mathbf{D}$; image restoration (IR) algorithms can be applied to obtain restored images, i.e., $\mathbf{R}$; and the human annotated quality score, i.e., mean opinion score (MOS) can be denoted as $s$. In the training phase, all of these $\{\mathbf{H},\mathbf{D},\mathbf{R},s\}$ are available. While in the testing phase, the model takes as input only $\{\mathbf{D},\mathbf{R}\}$ and predicts the quality score $s$.

\textbf{Formulation of CKDN.} CKDN consists of three modules, i.e., a degradation-tolerant embedding module (DTE), a quality-sensitive embedding module (QSE), and a convolutional score predictor (CSP), which are denoted as $\mathbf{\mathcal{E}_1}$, $\mathbf{\mathcal{E}_2}$, $\mathbf{\mathcal{S}}$, respectively. We can obtain the quality score by:
\begin{equation}\label{eqn:formulation}
s = \mathbf{\mathcal{S}}(\mathbf{\mathcal{E}}_1(\mathbf{D})-\mathbf{\mathcal{E}}_2(\mathbf{R})),
\end{equation}
where $\mathbf{D}$, $\mathbf{R}$, and $s$ are the degraded image, restored image, and quality score of the restored image, respectively.

\begin{figure}[!t]
    \centering
    \includegraphics[ width=1\linewidth]{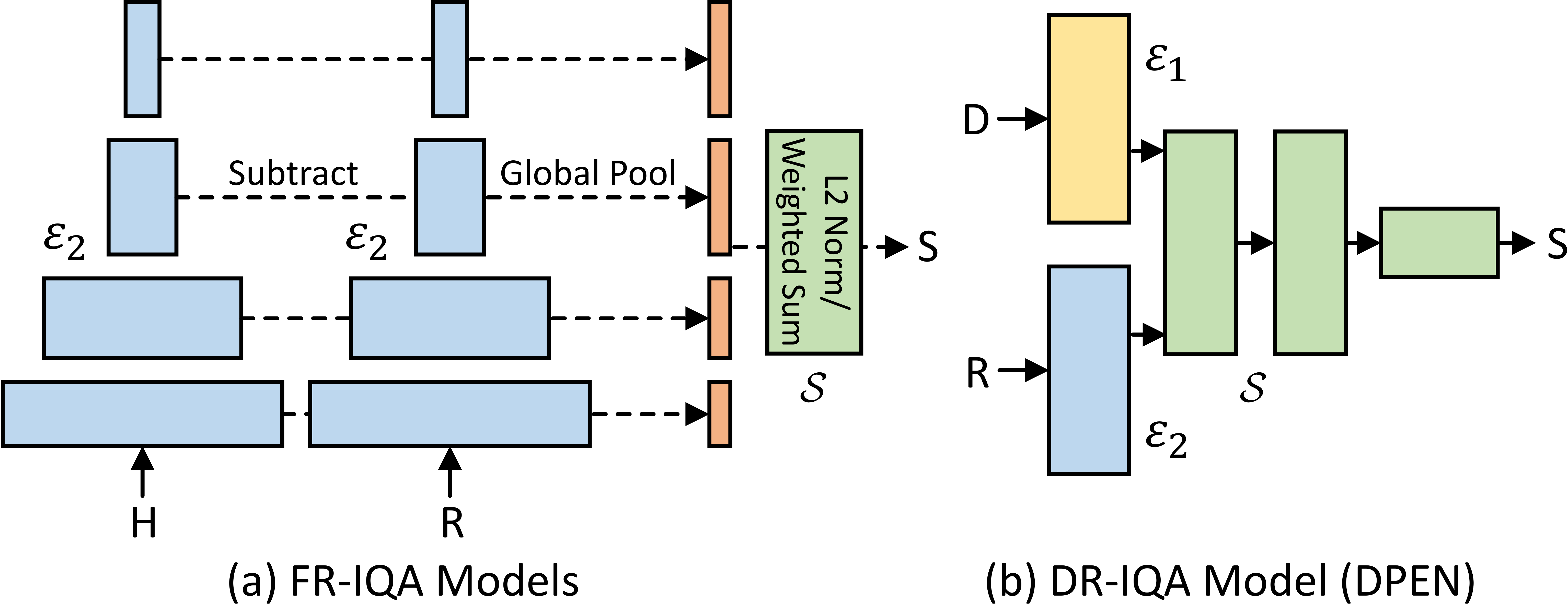}
    \vspace{-1mm}
    \caption{Comparison of (a) FR-IQA models and (b) DR-IQA model (CKDN). FR-IQA models focus on calculating difference between restored and reference images, while DR-IQA requires 1) different embeddings for restored and reference images and 2) convolutional layers for mapping feature difference to scores.}
    \vspace{-3mm}
    \label{fig:compare}
\end{figure}

\textbf{Special designs of CKDN for DR-IQA.} Compared to the state-of-the-art FR-IQA models~\cite{prashnani2018pieapp,zhang2018unreasonable}, the proposed CKDN has two advantages for the DR-IQA problem (as shown in Figure~\ref{fig:compare}). 1) FR-IQA models extract the restored image features and reference features with the same parameters, while our CKDN learns task-specific embeddings (i.e., the DTE and QSE). 2) FR-IQA models focus on extracting feature difference, while we pay more attention (by stacking residual blocks in the CSP) to mapping feature difference to quality scores. Such designs can better handle the degradation of reference images. 

\subsection{Learning quality-sensitive embedding}
Learning quality-sensitive features for restored images is important for image quality assessment. We propose to learn a quality-sensitive embedding (QSE) to extract such discriminative features. Regressing absolute quality scores is the most intuitive loss function to optimize the QSE, i.e., training the model to fit the input data $\{\mathbf{D},\mathbf{R}\}$ to the absolute quality score $s$. Specifically, we minimize the Mean Square Error (MSE) for such a regression problem:
\begin{equation}\label{eqn:La}
L_a = \frac{1}{N}\sum_{i}\norm{s_i - \mathbf{\mathcal{S}}(\mathbf{\mathcal{E}}_1(\mathbf{D})-\mathbf{\mathcal{E}}_2(\mathbf{R}_i))}_2^2,
\end{equation}
where $N$ is the number of training images. Note that in this paper, we show the case of single content with single degradation for simplicity.

\subsection{Learning degradation-tolerant embedding}
In this subsection, we will introduce how to learn the DTE, which is the key module of our model. the DTE is supposed to capture reference information from degraded images. Our basic idea is to learn reference information from pristine-quality images and use the learned knowledge to guide the learning of DTE. To achieve this, we propose a conditional knowledge distillation loss and design an effective pre-training method.

\textbf{Conditional knowledge distillation.} Knowledge distillation is powerful tool for feature learning~\cite{hinton2015distilling,yim2017gift,zhang2020object,zheng2019looking}. Our extended conditional knowledge distillation leverages restored images as context information (condition), which is important in various vision tasks~\cite{yu2018generative,zha2019context}. Equation~\ref{eqn:La} shows the basic objective function for learning CKDN. Let us consider training CKDN in a full-reference setting:
\begin{equation}\label{eqn:LaH}
L^H_a = \frac{1}{N}\sum_{i}\norm{s_i - \mathbf{\mathcal{S}}^H(\mathbf{\mathcal{E}}^H_1\mathbf{(H)}-\mathbf{\mathcal{E}}^H_2(\mathbf{R}_i))}_2^2,
\end{equation}
where $L^H_a$, $\mathbf{\mathcal{E}}^H_1(\cdot)$, $\mathbf{\mathcal{E}}^H_2(\cdot)$, and $\mathbf{\mathcal{S}}^H(\cdot)$ denote the score regression loss, degradation-tolerant embedding module, quality-sensitive embedding module, and the convolutional score predictor that trained with pristine-quality images. 

Compared to degraded images, it is much easier to learn deep image priors for IQA from pristine-quality images, i.e., $\mathbf{\mathcal{E}}^H_1\mathbf{(H)}$ is supposed to be better than $\mathbf{\mathcal{E}_1(D)}$. Such an observation motivates us to guide the learning of $\mathbf{\mathcal{E}_1(D)}$ by $\mathbf{\mathcal{E}}^H_1\mathbf{(H)}$. Specifically, our proposed conditional knowledge distillation loss can be denoted as:
\begin{equation}\label{eqn:CKD}
\begin{split} 
& L_{ckd} = L_a + L^H_a + \lambda \norm{\mathbf{\mathcal{E}}^H_1\mathbf{(H)}-\mathbf{\mathcal{E}}_1(\mathbf{D})}_2^2, \\
& s.t., \quad\mathbf{\mathcal{E}}^H_2=\mathbf{\mathcal{E}}_2, \quad\mathbf{\mathcal{S}}^H=\mathbf{\mathcal{S}},
\end{split}
\end{equation}
where $L_{ckd}$ is the conditional knowledge distillation loss. $L_{a}$, and $L^H_{a}$ are the score regression loss, which are shown in Equation~\ref{eqn:La} and Equation~\ref{eqn:LaH}, respectively. $\lambda$ is the loss weight.  $\mathbf{\mathcal{E}}^H_2=\mathbf{\mathcal{E}_2}$, and $\mathbf{\mathcal{S}}^H=\mathbf{\mathcal{S}}$ are the constraints. Note that such constraints are essential to achieve effective distillation for DR-IQA, which can ensure $\mathbf{\mathcal{E}_1(D)}$ and $\mathbf{\mathcal{E}}^H_1\mathbf{(H)}$ to be in the same latent space. In other word, the learning of embeddings for pristine-quality and degraded images should be in the same condition.

\textbf{Pre-training for conditional knowledge distillation.} 
The objective function in Equation~\ref{eqn:CKD} can be applied to train the CKDN in an end-to-end manner, where the parameters are initialized by ImageNet~\cite{krizhevsky2012imagenet} pre-trained parameters. Taking one step further, we study how the condition in Equation~\ref{eqn:CKD} influences the knowledge distillation performance. We find that pre-training the QSE, i.e., $\mathbf{\mathcal{E}}_2$, can advance the optimization and achieve better performance. That is, before training the CKDN by Equation~\ref{eqn:CKD}, we first learn the QSE by a relative score regression loss. Specifically, we train the model by comparing two restored images, e.g., $\mathbf{R_i},\mathbf{R_j}$, and regressing the possibility of $s_\mathbf{i}>s_\mathbf{j}$. Such loss can be denoted as:
\begin{equation}\label{eqn:Lr}
L_r = \frac{1}{N}\sum_{i,j}\norm{\text{Pr}{(s_i>s_j)} - \mathbf{\mathcal{S}}(\mathbf{\mathcal{E}}_2(\mathbf{R}_i)-\mathbf{\mathcal{E}}_2(\mathbf{R}_j))}_2^2,
\end{equation}
where $N$ is the number of training pairs, $\text{Pr}{(s_i>s_j)}$ can be obtained by $\text{Pr}{(s_i>s_j)} = 1/(1+10^{(s_j-s_i)/M})$~\cite{jinjin2020pipal}, $M$ is a dataset-specific parameter of the distribution, and $M=400$ in the dataset we used.

We call such training stage ``pre-training'' because the relative score regression loss enlarges the training data by creating pairs. And the pre-trained QSE can provide a better condition for the subsequent knowledge distillation.

\begin{table}[t]
\caption{The details of the subset of PIPAL~\cite{jinjin2020pipal} dataset that used in our experiments. \# IR indicates the number of IR models and settings, and \#Img means the number of distorted images.}
\label{tab:pipal}
\vspace{2mm}
\resizebox{1\columnwidth}{!}{\begin{tabular}{c|l|c|c}
\hline
IR tasks                                                       & Sub-tasks                                                                                                                      & \# IR & \# Img \\ \hline \hline
\begin{tabular}[c]{@{}c@{}c@{}c@{}}SR\end{tabular}   & \begin{tabular}[l]{@{}l@{}l@{}l@{}} (a) interpolation \\ (b) traditional SR\\ (c) PSNR-oriented SR \\ (d) GAN-based SR\end{tabular} & 62                        & 12,400                \\ \hline
Denoise                                                    & \begin{tabular}[l]{@{}l@{}l@{}} (e) mean filtering \\ (f) traditional denoising \\ (g) Deep Denoising\end{tabular}                    & 13                        & 2,600               \\ \hline
\begin{tabular}[c]{@{}c@{}c@{}}Mixture\end{tabular} &  \begin{tabular}[l]{@{}l@{}l@{}} (h) noisy images SR  \\ (i) SR after denoising \\ (j) SR after decompression\end{tabular}           & 14                    & 2,800               \\ \hline
\end{tabular}
}
\vspace{-6mm}
\end{table}

\textbf{Comparisons to previous work.} We analyze the difference between our proposed conditional knowledge distillation and previous arts, e.g., data augmentation, vanilla knowledge distillation~\cite{hinton2015distilling}, and hallucinated reference~\cite{lin2018hallucinated}.

The data augmentation method adds the $\{\mathbf{H},\mathbf{R_i}\}$ pair to training data, which would improve the generality of the model. Such a strategy can be regarded as an indirect knowledge distillation in a \textit{parameter-sharing} manner. 

The vanilla knowledge distillation method optimizes the model by Equation~\ref{eqn:CKD} without additional constraints. Such a strategy distills \textit{unaligned feature distributions} from the teacher network (trained with pristine-quality images) to the student network (trained with degraded images).

Hallucinated reference methods can be applied to DR-IQA by generating hallucinated references from the degraded images, and use the generated reference to conduct IQA. Such a solution tries to ``distill'' \textit{all the information} from pristine-quality images to degraded images.

Compared to data augmentation and vanilla knowledge distillation, our method is more direct and effective with well-aligned features. And compared to hallucinated reference methods, our selectively feature distillation is easier to learn as it is an ill-posed problem to recover all information from degraded images to pristine-quality images.

\section{Experiments}
\label{sec:exp}

\subsection{Experiment setup}
\label{sec:setup}

\textbf{Datasets.} We evaluate our model on the Perceptual Image Processing ALgorithms dataset (PIPAL)~\cite{jinjin2020pipal}, which is the largest IQA dataset with 1.13 million human annotations. PIPAL contains two types of images which are traditional distorted images (e.g., JPEG compression, noise, spatial warping, etc.) and images restored by IR algorithms. Since our proposed IQA method is designed for evaluating IR algorithms, we remove the traditional distorted images and use a subset of the released PIPAL, which contains three important image restoration tasks (i.e., super-resolution, de-noising, and mixture restoration). We split training and validation sets in a manner that is consistent with the real-world scenario for developing novel IR algorithms. Specifically, we randomly split 200 reference images into 175 training images and 25 validation images, which can ensure us to validate our model on unseen image contents. Moreover, we further split 89 IR algorithms into 50 algorithms for training and 39 algorithms for validation, which can ensure validating on unseen algorithms. Note that we split old algorithms for training and recent algorithms for validation. More details can be found in Table~\ref{tab:pipal}.

\textbf{Evaluation metrics.}
We follow previous works~\cite{bosse2017deep,jinjin2020pipal,ma2017end}, to use two most widely obtained IQA metrics, i.e., Spearman's Rank order Correlation Coefficients (SRCC)~\cite{sheikh2006statistical} and Pearson
Linear Correlation Coefficients (PLCC). SRCC evaluates the monotonicity of two lists, i.e., whether the order of predicted scores is consistent with human-annotated orders. And PLCC considers both orders and values.

\textbf{Baselines.}
We compared our method to previous methods in image quality assessment, including both NR-IQA and FR-IQA methods. In the following, the first four methods are NR-IQA models, and the latter five are FR-IQA models. We compare to them, due to their state-of-the-art performance and high relevance.

\begin{table}
{\small
\caption{Comparisons to previous works. We implement FR-IQA models to our DR-IQA setting, $\uparrow$ indicates the higher the better.}
\label{tab:compare}
\begin{center}
\resizebox{0.83\columnwidth}{!}{
    \begin{tabular}{c|c|c|c}
    \hline
    & Methods & SRCC$\uparrow$ & PLCC$\uparrow$\\
    \hline \hline
\multirow{4}{*}{NR-IQA}  
    & DIQaM~\cite{bosse2017deep} & 0.6255 & 0.6019 \\ 
    & NIMA~\cite{talebi2018nima} & 0.6330 & 0.6540 \\ 
    & Hall-IQA \cite{lin2018hallucinated} & 0.6390 & 0.6167 \\ 
    & MEON~\cite{ma2017end} & 0.6436 & 0.6610 \\ 
    \hline
\multirow{6}{*}{DR-IQA} &   PSNR & 0.3231 & 0.3516 \\
    & SSIM~\cite{wang2004image} & 0.3573 & 0.3509 \\
    & PieAPP~\cite{prashnani2018pieapp} & 0.6563 & 0.5937 \\
    & LPIPS~\cite{zhang2018unreasonable} & 0.6650 & 0.5872 \\
    & SWDN~\cite{jinjin2020pipal} & 0.6729 & 0.6052 \\
    \cline{2-4} 
  & CKDN (Ours) & \textbf{0.7669} & \textbf{0.7514} \\
    \hline
    \end{tabular}
    }
\end{center}
}
 \vspace{-4 mm}
\end{table}

\begin{itemize}[nosep]
\item  DIQaM \cite{bosse2017deep}: Propose to jointly learn the local quality and local weights, i.e., the relative importance of local quality to the global quality estimate.
\item  MEON \cite{ma2017end}: Formulate the NR-IQA to a multi-task training problem, i.e., a distortion identification task and a quality prediction task.
\item  NIMA \cite{talebi2018nima}: Propose to predict the distribution of MOS and introduced an EMD-based loss that penalizes misclassifications based on class distances.
\item  Hall-IQA \cite{lin2018hallucinated}: Propose to compare the distorted images to hallucinated references, which are restored from the distorted images by GAN.
\item  PSNR : Peak signal-to-noise ratio (PSNR) is an extension of mean square error (MSE), which further considers the ratio between the signal and noise.
\item  SSIM \cite{wang2004image}: The structural similarity index measure (SSIM) considers image degradation as a perceived change in structural information. 
\item  PieAPP \cite{prashnani2018pieapp}: Perceptual Image-Error Assessment through Pairwise Preference (PieAPP) proposed a pairwise-learning framework.
\item  LPIPS \cite{zhang2018unreasonable}: Learned Perceptual Image Patch Similarity (LPIPS) proposed to use a learned weighted sum of deep features as a perceptual metric.
\item  SWDN \cite{jinjin2020pipal}: Space Warping Difference IQA Network (SWDN) proposed to consider the robustness of spatial misalignment, especially for GAN-based distortions.
\end{itemize}

\textbf{Implementation details.}
We follow previous work~\cite{jinjin2020pipal} and set the input image resolution to $288 \times 288$. The learning rate is set to 0.15, and we use a constant learning rate strategy with a warm-up. We first train 10 epochs to initialize the quality-sensitive embedding, and then we train the CKDN by 20 epochs. we find the loss weight $\lambda$ in Equation~\ref{eqn:CKD} is robust to optimization, which is empirically set to 10. All the distortions are combined during training, and our model is trained end-to-end. The model size is 103MB and the training memory is 6.8GB (the batch size is 8 per GPU). We use PyTorch~\cite{paszke2019pytorch} as our codebase. Our CKDN is efficient, and each experiment can be finished in 30 minutes on 8 Tesla V100 GPUs.

\textbf{Architectures of CKDN.} We implement the two embeddings, i.e., the DTE and QSE, by a convolutional layer and three residual blocks~\cite{he2016deep}; and we implement the convolutional score predictor by four residual blocks and three fully connected layers. Note that although the DTE and QSE share the same architecture, their training strategies are different and specifically designed.

\begin{table}
{\small
\caption{Ablation experiments for main components.}
\vspace{2 mm}
\label{tab:ablation}
\begin{center}
\resizebox{0.75\columnwidth}{!}{
    \begin{tabular}{c|c|c}
    \hline
    Methods & SRCC$\uparrow$ & PLCC$\uparrow$\\
    \hline \hline
    FR model \cite{jinjin2020pipal} & 0.6729 & 0.6052 \\
    \hline
    CKDN &  0.7127 & 0.6606 \\
    CKDN + CKD &  0.7549 & 0.7473 \\
    CKDN + CKD + Pret. &  \textbf{0.7669} & \textbf{0.7514} \\
    \hline
    FR upper bound & \textbf{0.7922} & \textbf{0.8233} \\
    \hline
    \end{tabular}
    }
\end{center}
}
 \vspace{-3 mm}
\end{table}

\begin{table}
{\small
\caption{Ablation experiments on conv. blocks w/o CKD.}
\vspace{2 mm}
\label{tab:conv}
\begin{center}
\resizebox{0.92\columnwidth}{!}{
    \begin{tabular}{c|c|c|c}
    \hline
    Methods & Conv. blocks   & SRCC$\uparrow$ & PLCC$\uparrow$\\
    \hline \hline
    FR model \cite{jinjin2020pipal} & VGG & 0.6729 & 0.6052 \\
    \hline
    CKDN (Ours) & VGG & 0.7006 & 0.6420 \\
    CKDN (Ours) & Residual  & \textbf{0.7127} & \textbf{0.6606} \\
    \hline
    \end{tabular}
    }
\end{center}
}
 \vspace{-7 mm}
\end{table}

\subsection{Comparisons}

To evaluate our proposed CKDN, we implement the methods proposed for both NR-IQA and FR-IQA tasks. Specifically, for NR-IQA models, we only use the restored images for validation; and for FR-IQA models, we keep the same setting as our model and used the degraded images as references. We train and validate all the models on the same train/validation split to keep the comparison fair. The results can be found in Table~\ref{tab:compare}. It can be observed that our proposed CKDN outperforms all previous works with a clear margin, which yields the best solution when pristine-quality images are not available. PSNR and SSIM cannot work well with degraded references as the distance calculated by these methods are not robust for degradations. Moreover, without well-designed mechanisms for degraded reference, FR-IQA models cannot leverage such information and can only slightly outperform NR-IQA models in the term of SRCC. FR-IQA models even perform worse in the term of PLCC. Since PLCC is influenced by the value of scores, different reference images (e.g., $2\times$ and $4\times$ downsampled LR images) would cause disturbances and be considered as noises. However, with our proposed method, the degraded images can be leveraged to achieve 0.0940 SRCC gains compared to the state-of-the-art model~\cite{jinjin2020pipal}.

\begin{table}
{\small
\caption{Ablation experiments on model architectures.}
\vspace{2 mm}
\label{tab:archs}
\begin{center}
\resizebox{0.88\columnwidth}{!}{
    \begin{tabular}{c|c|c|c}
    \hline
    Embedding & Feature size & SRCC$\uparrow$ & PLCC$\uparrow$\\
    \hline \hline
    Shared & $72 \times 72$ & 0.7311 & 0.6926 \\
    Unshared & $36 \times 36$  & 0.7436 & 0.7387 \\
    Unshared & $18 \times 18$ & 0.7234 & 0.7176 \\
    Unshared & $72 \times 72$  & \textbf{0.7669} & \textbf{0.7514} \\
    \hline
    \end{tabular}
    }
\end{center}
}
 \vspace{-4 mm}
\end{table}

\begin{table}
{\small
\caption{Ablation experiments on knowledge distillations.}
\vspace{2 mm}
\label{tab:kd}
\begin{center}
\resizebox{0.96\columnwidth}{!}{
    \begin{tabular}{c|c|c}
    \hline
    Methods & SRCC$\uparrow$ & PLCC$\uparrow$\\
    \hline \hline
    w/o distillation & 0.7127 & 0.6606 \\
    Data augmentation &  0.7469 & 0.7087 \\ 
    Hallucinated reference &  0.7348 & 0.7087 \\
    Knowledge distillation &  0.7338 & 0.7093 \\
    \hline
    Conditional knowledge distillation &  \textbf{0.7669} & \textbf{0.7514} \\
    \hline
    \end{tabular}
    }
\end{center}
}
 \vspace{-7 mm}
\end{table}

\subsection{Ablation Studies}

\textbf{Main components.} We evaluate the main components of our proposed CKDN in Table~\ref{tab:ablation}. It can be observed that our proposed backbone works well for DR-IQA, which outperforms previous FR models by 0.0398 SRCC. Moreover, the proposed conditional knowledge distillation mechanism can further boost the performance by 0.0422 SRCC, which demonstrates the effectiveness of guiding DTE by pristine-quality images. Our proposed pre-training method can improve more than 0.01 SRCC, which makes our DR setting's performance close to the full reference upper bound. 

\textbf{Model architectures.} We study core techniques for model design in Table~\ref{tab:archs}. Sharing parameters for the quality-sensitive embedding and the degradation-tolerant embedding would cause 0.0358 SRCC drops, which show the necessity of learning unshared embeddings. Moreover, we want to emphasize that the spatial information of feature difference is important, which is ignored by FR-IQA models. Specifically, by down-sampling the input feature of CSP, we can observe 0.0233 SRCC drops. Last but not least, Table~\ref{tab:conv} shows that it is the formulation of CKDN that matters and the influence of specific architectures (e.g., residual or VGG~\cite{Simonyan15} blocks) are marginal.

\textbf{Knowledge distillation.} Table~\ref{tab:kd} shows the effectiveness of different knowledge distillation methods. As analyzed in Section~\ref{sec:method}, our proposed conditional knowledge distillation can more effectively distill useful information for quality assessment into the degradation-tolerant embedding. We find that hallucinated references even bring a performance drop, which would be caused by involving feature noises when restoring the hallucinated references. 

\textbf{Pre-training methods.} Table~\ref{tab:init} shows how different pre-training methods (pre-trained by different loss with different epochs) influence the performance. It can be observed that the proposed relative score regression loss performs better than the absolute score regression loss and ImageNet pre-training. Such improvements show the effectiveness of training on highly related tasks with enlarged pairwise data.

\begin{table}
{\small
\caption{Ablation experiments on pre-training in the term of SRCC. ``Pret.'' indicates pre-training.}
\label{tab:init}
\begin{center}
\resizebox{0.93\columnwidth}{!}{
    \begin{tabular}{c|c|c|c|c}
    \hline
    Epochs & 5 & 7  & 10 & 15\\
    \hline
    \hline
    ImageNet Pret. & 0.7549 & 0.7549  & 0.7549 & 0.7549 \\
    $L_a$ Pret. & 0.7561 & 0.7574  & 0.7536 & 0.7543 \\
    $L_r$ Pret. & 0.7587 & 0.7629  & \textbf{0.7669} & 0.7643 \\
    \hline
    \end{tabular}
    }
\end{center}
}
 \vspace{-6 mm}
\end{table}

\textbf{Robustness on train/validation splits.} We further explored how train/validation splits influence the effectiveness of the proposed methods. First, we study various train/validation splits from the aspect of image contents. We find that different train/validation splits lead to comparable performance (within 0.01 SRCC). Moreover, we study three different train/validation splits from the aspect of distortion types. The results can be found in Table~\ref{tab:trainval}, where the ``Baseline'' indicates SWDN~\cite{jinjin2020pipal}. It can be observed that our proposed method can achieve consistent improvements on all three settings, and are close to the performance achieved by using pristine-quality images.

\begin{table}
{\small
\caption{Extensive experiments for robustness evaluation by different train/validation splits (SRCC).}
\label{tab:trainval}
\begin{center}
\resizebox{0.94\columnwidth}{!}{
    \begin{tabular}{c|c|c|c}
    \hline
    Train/VAl Split & Baseline & Ours & FR setting \\
    \hline \hline
    50 train + 39 val & 0.6729 & 0.7669 & \textbf{0.7922} \\
    54 train + 35 val & 0.6884 & 0.7794 & \textbf{0.8011} \\
    59 train + 30 val & 0.6561 & 0.7697 & \textbf{0.7943} \\
    \hline
    \end{tabular}
    }
\end{center}
}
 \vspace{-6 mm}
\end{table}

\begin{figure*}[!t]
    \centering
    \vspace{-3mm}
    \includegraphics[ width=\linewidth]{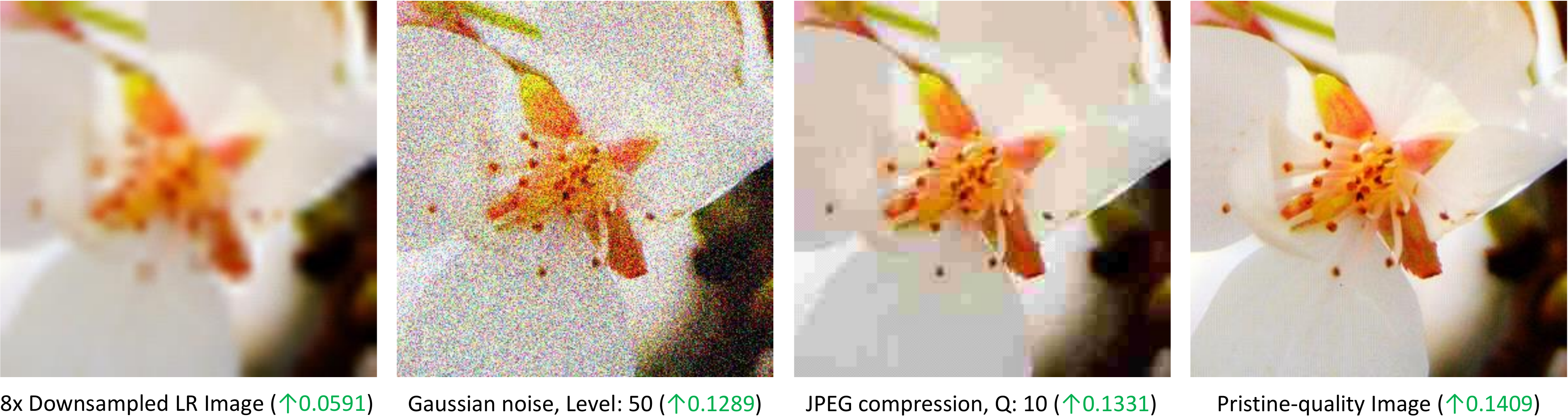}
    \vspace{-4mm}
    \caption{An illustration of the SRCC improvements obtained by different types of reference images. The percentage numbers marked in green is the SRCC improvements compared to no-reference settings.}
    \label{fig:case}
    \vspace{-5mm}
\end{figure*}

\textbf{CKDN can improve SR algorithms as a loss term}. Specifically, we follow the setting of NTIRE 2020 challenge winner~\cite{jo2020investigating}, i.e., use DIV8K~\cite{gu2019div8k} dataset and obtain LPIPS~\cite{zhang2018unreasonable} as a metric. The results (85k iterations) can be found in Table~\ref{tab:loss}. It can be observed that our CKDN can improve the SR performance. We only conduct experiments on the above settings, and more extensive experiments would be conducted in future work.

\begin{table}
{\small
\caption{CKDN can be used as a loss term. ``Baseline'' and ``VGG'' indicate ESRGAN~\cite{wang2018esrgan} and Perceptual Loss, respectively.}
\vspace{-2 mm}
\label{tab:loss}
\begin{center}
\resizebox{\columnwidth}{!}{
    \begin{tabular}{c|c|c|c}
    \hline
    Method & Baseline & Baseline + VGG & Baseline + CKDN \\
    \hline 
    LPIPS $\downarrow$ & 0.5380 & 0.5039 & \textbf{0.4957} \\
    \hline
    \end{tabular}
    }
\end{center}
}
 \vspace{-4 mm}
\end{table}

\textbf{CKDN can generalize well across datasets}. We test our model on an unseen dataset BAPPS~\cite{zhang2018unreasonable}, and Table~\ref{tab:cross} shows that our model can achieve comparable performance even with degraded references and without fine-tuning. 

\begin{table}
{\small
\caption{Experiments on cross dataset evaluation. 2AFC scores are the original metric of BAPPS~\cite{zhang2018unreasonable}.}
\vspace{-4 mm}
\label{tab:cross}
\begin{center}
\resizebox{\columnwidth}{!}{
    \begin{tabular}{c|c|c|c}
    \hline
    Reference & FR-IQA (LPIPS) & $\times$2 LR & $\times$4 LR  \\
    \hline 
    2AFC scores $\uparrow$ & \textbf{69.80} & 69.71 & 68.98  \\
    \hline \hline
    Reference & $\times$8 LR & L25 noise  & L50 noise \\
    \hline 
    2AFC scores $\uparrow$ & 68.38 & 68.58 & 67.63  \\
    \hline
    \end{tabular}
    }
\end{center}
}
 \vspace{-4 mm}
\end{table}

\textbf{CKDN can help in practice}. In Table~\ref{tab:prac}, we replace subjective image quality experiments from the state-of-the-art SPSR~\cite{ma2020structure}. The baseline is SRGAN~\cite{ledig2017photo}. CKDN delivers the same conclusion as PI~\cite{blau2018perception} and LPIPS~\cite{zhang2018unreasonable}. We also find some inconsistent results, when evaluating comparable IR algorithms. And this problem would be further studied in the future.

\begin{table}
{\small
\caption{ The results in practice: use CKDN as a SR metric. The results are in the format of PI$\downarrow$/LPIPS$\downarrow$/(1-CKDN)$\downarrow$.}
\vspace{-4 mm}
\label{tab:prac}
\begin{center}
\resizebox{\columnwidth}{!}{
    \begin{tabular}{c|c|c|c}
    \hline
    Method & Set5 & Set14 & BSD100  \\
    \hline 
    Bicubic & 7.36/0.34/0.48 & 7.02/0.43/0.41 & 7.00/0.52/0.51  \\
    \hline 
    SRGAN & 3.98/0.08/0.40 & 3.08/0.16/0.35  & 2.54/0.19/0.43 \\
    \hline 
    SPSR & \textbf{3.27/0.06/0.36} & \textbf{2.90/0.13/0.32} & \textbf{2.35/0.16/0.40}  \\
    \hline
    \end{tabular}
    }
\end{center}
}
 \vspace{-6 mm}
\end{table}

\vspace{-2mm}

\section{Discussions}
\label{sec:dis}

In this section, we conduct extensive experiments and detailed analyses to 1) evaluate the robustness of CKDN, 2) show the extreme effectiveness of reference images for evaluating images generated by GANs, and 3) provide an intuitive comprehension of how well current IQA models can be used as IR metrics.

\textbf{The proposed method works well on degraded references with a large range of degradation types.} We proposed to use IR inputs, i.e., degraded images, to help to assess the quality of restored images, which is practical in real-world scenarios. We further studied using the same degraded image (e.g., $4\times$ downsampled LR image) as a reference for assessing all the restored (e.g., SR, Denoised) images. Such a setting is not practical, while it can intuitively show the influence of reference quality on IQA performance. The results can be found in Table~\ref{tab:robust} and an illustration can be found in Figure~\ref{fig:case}. We choose the most common 5 degradations in the dataset. It can be observed that the proposed method performs well with a large range of degradation, e.g., $2\times$ downsample, $4\times$ downsample, Gaussian noise with levels: 25 and 50, JPEG compression with quality of 30 and 10. The performance dropped on the $8\times$ downsample setting, while still outperforms the no-reference setting by 0.0591 SRCC.

\textbf{Reference images is extremely important for evaluating images generated by GANs.} Evaluating GAN-based IR models attracts lots of attention in recent years~\cite{gu2020giqa,jinjin2020pipal,zhang2018unreasonable} as the artifacts generated by GANs yield new challenges for IQA. We study the performance of our model on GAN-based images in Table~\ref{tab:acc} and find introducing reference (e.g., full reference or degraded reference) can significantly benefit assessing GAN-based images (with 0.3101 SRCC gains). Such a result is reasonable as the textures generated by GANs may confuse NR-IQA models on distinguishing the artifacts and image contents. Thus reference images can provide strong guidance to disentangle quality-sensitive factors from image contents.

\begin{table}
{\small
\caption{Extensive experiments on how the quality of reference images influence IQA performances.}
\label{tab:robust}
\begin{center}
\resizebox{0.88\columnwidth}{!}{
    \begin{tabular}{c|c|c|c}
    \hline
    \multicolumn{2}{c|}{Reference} & SRCC$\uparrow$ & PLCC$\uparrow$\\
    \hline \hline
    \multicolumn{2}{c|}{NR baseline} & 0.6513 & 0.6312 \\
    \hline
    \multirow{3}{*}{LR images}&$2 \times$ &  0.7859 & 0.8286 \\
    &$4 \times$ &  0.7639 & 0.7806 \\
    &$8 \times$ &  0.7104 & 0.7251 \\ \cline{1-4}
    \multirow{2}{*}{Gauss. noise}&Level: 25 &  0.7888 & 0.8291 \\
    &Level: 50 &  0.7802 & 0.8190 \\ \cline{1-4}
    \multirow{2}{*}{JPEG}& Q: 30 & 0.7911 & 0.8239 \\
    &Q: 10 & 0.7844 & 0.8288 \\ \cline{1-2}
    \hline
    \multicolumn{2}{c|}{IR inputs (our DR setting)} & 0.7669 & 0.7514 \\
    \hline
    \multicolumn{2}{c|}{FR upper bound} & \textbf{0.7922} & \textbf{0.8233} \\
    \hline
    \end{tabular}
    }
\end{center}
}
 \vspace{-9 mm}
\end{table}

\textbf{Current IQA methods can provide around 85\% reliable judgments when they are adopted as IR metrics.} We provide an intuitive comprehension of how well current IQA models can be used as IR metrics. In particular, we use IQA models to compare IR algorithms under the same task and setting (e.g., the $\times 4$ SR task by GAN) and calculate the accuracy of the predicted judgments. We train all the settings by the same backbone, i.e., our CKDN, for fair comparisons. The results in Table~\ref{tab:acc} show that the accuracy of the FR-IQA setting is around 85\%, which means 85\% preferences of the IQA model are consistent with human judgments. Moreover, DR-IQA can achieve comparable results with the FR-IQA setting, while the NR-IQA setting causes 17.27\% accuracy drops.

\begin{table}
{\small
\caption{Extensive experiments on evaluating GAN-based images by SRCC and accuracy. Accuracy: we use IQA models to compare the output images of two IR algorithms (which one is better), and calculates the percentage of right judgments (the same as human judgments). $\uparrow$ indicates the higher the better.}
\label{tab:acc}
\begin{center}
\resizebox{0.90\columnwidth}{!}{
    \begin{tabular}{c|c|c|c}
    \hline
    Metrics & NR-IQA & DR-IQA & FR-IQA \\
    \hline \hline
    SRCC$\uparrow$ & 0.4081 & 0.7182 & \textbf{0.7560} \\
    Accuracy $\uparrow$ &  0.6853  & 0.8507 & \textbf{0.8580}\\
    \hline
    \end{tabular}
    }
\end{center}
}
 \vspace{-9 mm}
\end{table}

\vspace{-2mm}
\section{Conclusion}
\vspace{-2mm}
\label{sec:con}
In this paper, we study the problem of evaluating IR models without pristine-quality images. We present a degraded-reference solution (DR-IQA) that introduces IR inputs to boost the quality assessment performance. we find it is nontrivial to leverage degraded images as reference, and by learning a Conditional Knowledge Distillation Network (CKDN), promising performance can be achieved. We have conducted extensive evaluations to demonstrate the effectiveness of our proposed method. Moreover, we draw three valuable conclusions about IQA and IR metrics: 1) our CKDN works well with a large range of degradation types, 2) reference images are extremely important for evaluating GAN-based images, and 3) current IQA methods can provide around 85\% reliable judgments for evaluating IR algorithms. In the future, we will focus on 1) extending our model to evaluate other conditional GAN tasks (e.g., in-painting, and semantic segmentation to images), and 2) further studying the problem of using our model as a loss/metric for IR algorithms.

\vspace{-3mm}
\section{Acknowledgement}
\vspace{-2mm}

This work was supported by the National Key R\&D Program of China under Grand 2020AAA0105702, National Natural Science Foundation of China (NSFC) under Grants U19B2038, the University Synergy Innovation Program of Anhui Province under Grants GXXT-2019-025.

\end{document}